\documentclass[10pt,english]{article}
\usepackage{mathptmx}
\usepackage[T1]{fontenc}
\usepackage[latin9]{luainputenc}
\usepackage{babel}
\usepackage{amsmath}
\usepackage{amsthm}
\usepackage{amssymb}
\usepackage{setspace}
\usepackage[pdfusetitle,
 bookmarks=true,bookmarksnumbered=false,bookmarksopen=false,
 breaklinks=false,pdfborder={0 0 1},backref=false,colorlinks=false]
 {hyperref}
\begin{document}
\title{On the thermodynamic limit of bipartite spin networks}
\author{Simone Franchini{\normalsize\thanks{Correspondence: simone.franchini@yahoo.it}\thanks{Sapienza Università di Roma, Piazza Aldo Moro 1, 00185 Roma, Italy}}}
\date{~}
\maketitle
\begin{abstract}
We investigate the properties of the thermodynamic limit in a general
bipartite spin network with pairwise interactions. This is done by
integrating one of the the spin groups, to transform the bipartite
problem into a single group problem with a non--linear Hamiltonian.
The transformed model is also relevant due to a similarity with the
LCVAE architecture. 

~

\noindent\textit{keywords: bipartite networks, Boltzmann machines,
neural networks, variational auto--encoders}
\end{abstract}
\newpage{}

\noindent In this paper we investigate the nature of the thermodynamic
limit in a general model of bipartite spin network with pairwise interactions
by transforming the bipartite problem into a single group problem
with non--linear Hamiltonian. Our method allows to study the thermodynamic
limit for the Asymmetric Bipartite Sherrington--Kirkpatrick model
(ABSK) \cite{Harnett,BipartiteBarra,Pak,Murrat_Bipartite,Chen_Hong_Bin,BSKold}
in straightforward way. The notation is that of \cite{Franchini2024}.

\section{Definitions}

\noindent Let us call $A$ and $B$ the vertices sets of the first
and second spin group, respectively, the group sizes are: $|A|=\alpha N$
and $|B|=\bar{\alpha}N$, where $\alpha+\bar{\alpha}=1$, so that
the total number of spins is ultimately $N$. Then, the generic bipartite
Hamiltonian is as follows:
\begin{equation}
H\left(\sigma_{A},\sigma_{B}\right):=\sum_{i\in A}\sum_{j\in B}H_{ij}\sigma_{i}\sigma_{j}
\end{equation}
The associated canonical partition function is:
\begin{equation}
Z:=\sum_{\sigma_{A}\in\Omega_{A}}\sum_{\sigma_{B}\in\Omega_{B}}\exp\left(H\left(\sigma_{A},\sigma_{B}\right)\right)
\end{equation}
In this paper we will be especially interested in the Thermodynamic
Limit (TL), that is, the scaling limit of the Free Energy per spin.
The existence of such limit is an important problem in mathematical
Spin Glass theory \cite{BipartiteBarra,Pak,Murrat_Bipartite,Chen_Hong_Bin,BSKold}. 

Departing from the conventional approaches, we start by rewriting
the Hamiltonian in terms of the cavity fields that mediate the influence
of the group $B$ on the group $A$:
\begin{equation}
h_{i}\left(\sigma_{B}\right):=\sum_{j\in B}H_{ij}\sigma_{j}
\end{equation}
Introducing this new symbol we can rewrite the Hamiltonian as scalar
product between the magnetization state of group $A$ and the cavity
fields incoming from the other group $B$. Explicitly, the Hamiltonian
is rewritten as follows \cite{BSKold,Franchini2024}:
\begin{equation}
H\left(\sigma_{A},\sigma_{B}\right)=\sum_{i\in A}\sigma_{i}\,h_{i}\left(\sigma_{B}\right)
\end{equation}
From this reformulation we can see that it is possible to reduce the
number of spins to deal with by integrating over group $A$. Introducing
the abbreviations 
\begin{equation}
f\left(h_{i}\right):=\log2\cosh\left(h_{i}\right),\ \ \ F\left(h_{A}\right):=\sum_{i\in A}f\left(h_{i}\right)
\end{equation}
and after few manipulation we easily arrive to a new formula
\begin{equation}
Z=\sum_{\sigma_{B}\in\Omega_{B}}\exp\left(F\left(h_{A}\left(\sigma_{B}\right)\right)\right).
\end{equation}
This representation for the bipartite partition function that we first
used in the preprint \cite{BSKold} is such that the new Hamiltonian,
$F$, only depends on the spins of the second group through the cavity
fields (actually their distribution). Notice also that the new Hamiltonian
is non--linear. and it has been used already as a loss function to
improve variational auto--encoders performances (LCVAE, \cite{Chen,XuVAC}).

\section{Support of the cavity fields.}

From the last equation, we found that the partition function depends
on group $B$ trough the cavity fields, therefore the $|B|-$dimensional
binary space of magnetisations can be ``pushforwarded'' to the $|A|-$dimensional
real space of the cavity fields. To better study this property let
us introduce a symbol for the support of the cavity fields 
\begin{equation}
\Phi_{A}:=\bigcup_{\sigma_{B}\in\Omega_{B}}\left\{ {\scriptscriptstyle \,}h_{A}\left(\sigma_{B}\right)\right\} 
\end{equation}
that is simply the collection of all possible values that can be assumed
by the vector of the cavity fields as the input states varies. Using
this new ensemble we can rewrite the partition function as follows:
\begin{equation}
Z=\sum_{h_{A}\in\Phi_{A}}\ \exp\left(F\left(h_{A}\right)\right)
\end{equation}
In general, the cardinality (or complexity) of this set is smaller
than the input space due to possible congruence in the mapping of
different spin states
\begin{equation}
|\,\Phi_{A}|\leq|\,\Omega_{B}|=2^{|B|}
\end{equation}
This is the general inequality, but notice also that in many interesting
special cases, e.g., random couplings extracted according to some
continuous distributions, like the Gaussian couplings for the ABSK
model, any congruence is an event with zero probability measure and
therefore the bound would be tight almost surely.

\section{Eigenstates of magnetization}

Let us now introduce a suitable notation to deal with the magnetization
eigenstates. We define magnetization function as follows:
\begin{equation}
M\left(\sigma_{B}\right):=\sum_{i\in B}\sigma_{B}
\end{equation}
We define the magnetization eigenset as the subsets of spin states
that returns the same value of the total magnetization
\begin{equation}
\Omega_{B}\left(M\right):=\left\{ \sigma_{V}\in\Omega_{B}:\,M\left(\sigma_{B}\right)=M\right\} 
\end{equation}
This ensemble can be studied even at the ``sample--path'' level,
both at finite size and in the thermodynamic limit, through established
Large--Deviations (LD) techniques \cite{Dembo}. For example, the
asymptotic complexity can be computed exactly \cite{RSBwR,FranchiniSPA2023,FranchiniURNS2017,FranchiniBalzanIRT2023}
\begin{equation}
|\,\Omega_{B}\left(M\right)|\approx\exp\left(|B|\,\rho\left(M/|B|\right)\right)\label{eq:scaly}
\end{equation}
where $\rho$ is the scaling limit of the entropy density of a binary
string 
\begin{equation}
\rho\left(m\right)=-\log2+\frac{1}{2}\log\left(1-m^{2}\right)+\frac{m}{2}\log\left(\frac{1+m}{1-m}\right)
\end{equation}
Notice that if we flip one spin the total magnetization will change
by two units, therefore we have to pay attention when defining the
support of the total magnetization: 
\begin{equation}
\Gamma_{B}:=\left\{ 2k\in\mathbb{Z}:\,|k|\leq|B|/2\right\} 
\end{equation}
If we join the magnetization eigensets we recover the full spin support
\begin{equation}
\Omega_{B}=\bigcup_{M\in\Gamma_{B}}\Omega_{B}\left(M\right)
\end{equation}
therefore we can decompose the sum over the magnetization states into
a sum over the eigenstates and then a sum over eigenvalues
\begin{equation}
\sum_{\sigma_{B}\in\Omega_{B}}\left(\,\cdot\,\right)=\sum_{M\in\Gamma_{B}}\ \sum_{\sigma_{B}\in\Omega_{B}\left(M\right)}\left(\,\cdot\,\right)
\end{equation}
Therefore, the formula for the partition function becomes
\begin{equation}
Z=\sum_{M\in\Gamma_{B}}Z\left(M\right)
\end{equation}
where we introduced the restricted partition function
\begin{equation}
Z\left(M\right):=\sum_{\sigma_{B}\in\Omega_{B}\left(M\right)}\ \exp\left(F\left(h_{A}\left(\sigma_{B}\right)\right)\right)
\end{equation}
that is, in fact, the partition function restricted to a given eigenset. 

\section{Partition of the cavity space}

Also, the partitions of the spin space into the eigenstates of magnetization
can be push--forwarded to the space of the cavity fields, this is
done by introducing 
\begin{equation}
\Phi_{A}\left(M\right):=\bigcup_{\sigma_{B}\in\Omega_{B}\left(M\right)}\left\{ {\scriptscriptstyle \,}h_{A}\left(\sigma_{B}\right)\right\} 
\end{equation}
that is the projection of the eigenset on the dual cavity space. As
before, if we joint the projected eigensets we get back the full support
of the cavity fields
\begin{equation}
\Phi_{A}=\bigcup_{M\in\Gamma_{B}}\Phi_{A}\left(M\right)
\end{equation}
As noticed before, in general we have the following complexity bound
in place:
\begin{equation}
|\,\Phi_{A}\left(M\right)|\leq|\,\Omega_{B}\left(M\right)|
\end{equation}
The formula for the restricted partitions is simplified as follows
\begin{equation}
Z\left(M\right):=\sum_{h_{A}\in\Phi_{A}\left(M\right)}\ \exp\left(F\left(h_{A}\right)\right)
\end{equation}
It is therefore crucial to study the properties of the dual of the
support in cavity space, especially from a probabilistic point of
view.

\section{Eigendistributions }

We could characterize the magnetization ensembles at micro--canonical
level by introducing the eigenstates distribution
\begin{equation}
\zeta_{M}\left(\sigma_{B}\right):=\frac{1}{|\,\Omega_{B}\left(M\right)|}\,\mathbb{I}\left(\sigma_{B}\in\Omega_{B}\left(M\right)\right)
\end{equation}
that is just a uniform distribution over a magnetization ensemble
(with given magnetization). Let introduce a notation for the average
respect to the eigenstates 
\begin{equation}
\langle\mathcal{O}\left(\sigma_{B}\right)\rangle_{\zeta_{M}}:=\sum_{\sigma_{B}\in\Omega_{B}\left(M\right)}\zeta_{M}\left(\sigma_{B}\right)\mathcal{O}\left(\sigma_{B}\right)
\end{equation}
then we can rewrite the restricted partition functions once again
\begin{equation}
Z\left(M\right)=|\,\Omega_{B}\left(M\right)|\langle\,\exp\left(F\left(h_{A}\left(\sigma_{B}\right)\right)\right)\rangle_{\zeta_{M}}
\end{equation}
so that we have to compute the thermodynamic limit only for some given
magnetization ensemble. As we shall see in the next Section, this
allows to connect with the Random Energy Model (REM) \cite{Kurkova1,Kistler}.
Before going forward, it will be useful to indulge more on the formalism
and introduce a notation also for the push--forward measure:

\begin{equation}
\xi\left(h_{A}\right):=\frac{1}{|\,\Omega_{B}|}\sum_{\sigma_{B}\in\Omega_{B}}\delta^{\,\,\left(A\right)}\left(h_{A}-h_{A}\left(\sigma_{B}\right)\right)
\end{equation}
where $\delta^{\,\,\left(A\right)}$ is our notation for the Dirac
delta function in the cavity space. Similarly as before, the notation
for the average is as follows:
\begin{equation}
\langle\mathcal{O}\left(h_{A}\right)\rangle_{\xi}:=\int_{h_{A}\in\mathbb{R}^{A}}\mathcal{D}h_{A}\,\xi\left(h_{A}\right)\mathcal{O}\left(h_{A}\right),
\end{equation}
where the path--integral is defined as usual:
\begin{equation}
\int_{h_{A}\in\mathbb{R}^{A}}\mathcal{D}h_{A}:=\prod_{i\in A}\int_{h_{i}\in\mathbb{R}}dh_{i}
\end{equation}
The partition function is therefore
\begin{equation}
Z=|\,\Phi_{A}|\langle\,\exp\left(F\left(h_{A}\right)\right)\rangle_{\xi}\label{eq:frez}
\end{equation}
Introduce a notation for the push--forward of the magnetization eigensets:
\begin{equation}
\xi_{M}\left(h_{A}\right):=\frac{1}{|\,\Omega_{B}\left(M\right)|}\sum_{\sigma_{B}\in\Omega_{B}\left(M\right)}\delta^{\,\,\left(A\right)}\left(h_{A}-h_{A}\left(\sigma_{B}\right)\right)
\end{equation}
The restricted partition function is again
\begin{equation}
Z\left(M\right)=|\,\Phi_{A}\left(M\right)|\left\langle \,\exp\left(F\left(h_{A}\right)\right)\right\rangle _{\xi_{M}}\label{eq:saboom}
\end{equation}
The full partition function is the sum of the restricted partition
function on the possible values of total magnetization.

\section{Gaussian fields and REM universality}

We are now applying our considerations to a more specific random model,
and study the case of Gaussian couplings, that is, the Asymmetric,
Bipartite Sherrington--Krikpatrick model (ABSK, \cite{Harnett,BipartiteBarra,Pak,Murrat_Bipartite,Chen_Hong_Bin,BSKold}).
Hereafter
\begin{equation}
H_{ij}=\beta J_{ij}/\sqrt{N},\ \ \ J_{ij}\sim\gamma,\ \ \ \gamma\left(J_{ij}\right):=\frac{1}{\sqrt{2\pi}}\exp\left(-J_{ij}^{2}/2\right)
\end{equation}
where the couplings are i.i.d. Gaussian (actually Normal) random variables.
Following \cite{RSBwR,FranchiniSPA2023}, the first step is to separate
the fluctuations from their average. Introducing the fundamental parameters
\begin{equation}
\psi_{i}:=\frac{1}{|B|}\sum_{j\in B}H_{ij},\ \ \ \delta_{i}^{2}:=\frac{1}{|B|}\sum_{j\in B}H_{ij}^{2}-\psi_{i}^{2}
\end{equation}
we express the noise of the couplings using the normalized variable
\begin{equation}
\epsilon_{ij}:=\frac{H_{ij}-\psi_{i}}{\delta_{i}}
\end{equation}
Now, by introducing the vertex sub--set
\begin{equation}
V\left(\sigma_{B}\right):=\left\{ j\in B:\,\sigma_{j}=-1\right\} 
\end{equation}
that tracks which spins is flipped in a given state, and the auxiliary
parameter 
\begin{equation}
E:=\frac{1+M}{2}
\end{equation}
we can re--write the eigenset of magnetization as follows
\begin{equation}
\Omega_{B}\left(M\right):=\left\{ \sigma_{B}\in\Omega_{B}:\,|\,V\left(\sigma_{B}\right)|=E\right\} 
\end{equation}
In \cite{RSBwR,FranchiniSPA2023} it is shown that whenever
\begin{equation}
\sigma_{B}\in\Omega_{B}\left(M\right)
\end{equation}
we can define the crucial variable
\begin{equation}
\epsilon_{i}\left(\sigma_{B}\right):=\frac{1}{\sqrt{E}}\sum_{j\in R\left(\sigma_{B}\right)}\epsilon_{ij}
\end{equation}
Notice that $\epsilon_{i}$ is normalized. We can express the cavity
field as follows:
\begin{equation}
h_{i}\left(\sigma_{B}\right)=X_{i}+\epsilon_{i}\left(\sigma_{B}\right)\sqrt{Y_{i}}
\end{equation}
where the coefficients are given by
\begin{equation}
X_{i}=\psi_{i}\,M,\ \ \ Y_{i}=2\delta_{i}^{2}\left(\bar{\alpha}N-M\right)
\end{equation}
Most important, in Section 6.3 of \cite{FranchiniSPA2023} is shown
that is possible to further re--normalize the cavity field and express
it in term of a field with zero overlap matrix. We introduce
\begin{equation}
V'_{E}\left(\sigma_{B}\right):=\left\{ j\in B:\,\sigma_{j}\tau_{j}\left(E\right)=-1\right\} 
\end{equation}
where $\tau$ is the the ``target'' state (see Fig. 1 and Eq. (69)
of \cite{FranchiniSPA2023}) defined as follows:
\begin{equation}
\tau_{j}\left(E\right):=1-2\,\mathbb{I}\left(j\le E\right)
\end{equation}
and the size of the re--normalized set is
\begin{equation}
E':=\frac{1+M'}{2}
\end{equation}
Then, we can further split the eigenset into a finer partition
\begin{equation}
\Omega_{B}\left(\,M,\,M'\right):=\left\{ \sigma_{B}\in\Omega_{B}\left(M\right):\,|\,V'_{E}\left(\sigma_{B}\right)|=E-E'\right\} 
\end{equation}
As before we have to introduce the span of $M'$
\begin{equation}
\Gamma_{B}'\left(M\right):=\left\{ 2k\in\mathbb{Z}:\,|k|\leq E/2\right\} 
\end{equation}
then any sum respect to the support can be decomposed as follows:
\begin{equation}
\sum_{\sigma_{B}\in\Omega_{B}}\left(\,\cdot\,\right)=\sum_{M\in\Gamma_{B}}\ \sum_{M'\in\Gamma_{B}'\left(M\right)}\,\sum_{\sigma_{B}\in\Omega_{B}\left(M,\,M'\right)}\left(\,\cdot\,\right)
\end{equation}
The push--forward measure in cavity space is 
\begin{equation}
\xi_{M,\,M'}\left(h_{A}\right):=\frac{1}{|\,\Omega_{B}\left(\,M,\,M'\right)|}\sum_{\sigma_{B}\in\Omega_{B}\left(M,\,M'\right)}\delta^{\,\,\left(A\right)}\left(h_{A}-h_{A}\left(\sigma_{B}\right)\right)
\end{equation}
the support of the cavity fields is
\begin{equation}
\Phi_{A}\left(\,M,\,M'\right):=\bigcup_{\sigma_{B}\in\Omega_{B}\left(M,\,M'\right)}\left\{ {\scriptscriptstyle \,}h_{A}\left(\sigma_{B}\right)\right\} 
\end{equation}
and the partition function is 
\begin{equation}
Z=\sum_{M\in\Gamma_{B}}\ \sum_{M'\in\Gamma_{B}'\left(M\right)}\left|\,\Phi_{A}\left(\,M,\,M'\right)\right|\left\langle \,\exp\left(F\left(h_{A}\right)\right)\right\rangle _{\xi_{M,\,M'}}
\end{equation}
In Section 6.3 of \cite{FranchiniSPA2023} is shown that if we pick
the state from
\begin{equation}
\sigma_{B}\in\Omega_{B}\left(\,M,\,M'\right)
\end{equation}
we can express the cavity field as follows:
\begin{equation}
h_{i}\left(\sigma_{B}\right)=X_{i}+\epsilon_{i}'\left(\sigma_{B}\right){\textstyle \sqrt{Y_{i}'}}\label{eq:REMfild}
\end{equation}
where $\epsilon_{i}'$ is a field with zero overlap in distribution
and the new coefficient is
\begin{equation}
Y_{i}':=\left(1-E'/E\right)Y_{i}
\end{equation}
Crucially if the couplings are Gaussian then $\epsilon_{i}'$ is ultimately
a Gaussian field with a zero overlap matrix: this is just another
definition for the Random Energy Model (REM) of the Derrida's type.
Hence we found that $\Phi_{A}$ is a mixture of REMs: the existence
of the thermodynamic limit is ultimately deduced from well established
techniques from the literature on REM and its generalizations \cite{Kurkova1,Kistler,Arous-Kupsov}. 

\section{Ito's Lemma and Thermodynamic Limit}

In this section we will sketch how to actually deal with the Thermodynamic
limit of the ABSK model in the kernels framework (see also \cite{Franchini2024}).
We will start from Eq. (\ref{eq:REMfild}). Notice that, since the
couplings are normalized by $\sqrt{N}$, then the coefficients and
variables appearing in this equation are all of $O\left(1\right)$.
Let us introduce the following abbreviations for the first and second
derivative of $f$
\begin{equation}
\dot{f}\left(h_{i}\right):=\tanh\left(h_{i}\right),\ \ \ \ddot{f}\left(h_{i}\right):=1-\tanh\left(h_{i}\right)^{2}
\end{equation}
Then, by Taylor's theorem applied to $f$ we find
\begin{equation}
f\left(h_{i}\left(\sigma_{B}\right)\right)=f\left(X_{i}\right)+\epsilon_{i}'\left(\sigma_{B}\right)\sqrt{Y_{i}'}\dot{f}\left(X_{i}\right)+\frac{1}{2}\,\epsilon_{i}'\left(\sigma_{B}\right)^{2}Y_{i}'\ddot{f}\left(X_{i}\right)+\ ...
\end{equation}
By standard arguments it is possible to conclude that under Gibbs
average
\begin{equation}
\sum_{i\in A}\epsilon_{i}'\left(\sigma_{B}\right)^{2}Y_{i}'\ddot{f}\left(X_{i}\right)\overset{d}{=}\sum_{i\in A}Y_{i}'\ddot{f}\left(X_{i}\right)
\end{equation}
while the Gaussian sum rule ensures that
\begin{equation}
\sum_{i\in A}\epsilon_{i}'\left(\sigma_{B}\right)\sqrt{Y_{i}'}\dot{f}\left(X_{i}\right)\overset{d}{=}\epsilon'\left(\sigma_{B}\right)\sqrt{\sum_{i\in A}Y_{i}'\dot{f}\left(X_{i}\right)^{2}}
\end{equation}
where $\epsilon'$ is again a REM of the Derrida type. We obtained
the Ito's lemma:
\begin{equation}
F\left(h_{A}\left(\sigma_{B}\right)\right)\overset{d}{=}F\left(X_{A}\right)+\frac{1}{2}\sum_{i\in A}{\textstyle Y_{i}'}\ddot{f}\left(X_{i}\right)+\epsilon'\left(\sigma_{B}\right)\sqrt{\sum_{i\in A}Y_{i}'\dot{f}\left(X_{i}\right)^{2}}
\end{equation}
Introducing the abbreviations
\begin{equation}
F':=F\left(X_{A}\right)+\frac{1}{2}\sum_{i\in A}Y_{i}'\ddot{f}\left(X_{i}\right),\ \ \ K':=\sum_{i\in A}Y_{i}'\dot{f}\left(X_{i}\right)^{2}
\end{equation}
we can rewrite the Ito formula in the following compact form
\begin{equation}
F\left(h_{A}\left(\sigma_{B}\right)\right)\overset{d}{=}F'+\epsilon'\left(\sigma_{B}\right)\sqrt{K'}
\end{equation}
Substituting in the formulas of the previous section we find:
\begin{equation}
\left\langle \,\exp\left(F\left(h_{A}\right)\right)\right\rangle _{\xi_{M,\,M'}}=\exp{\textstyle \left(F'\right)}\,\langle\,\exp\,(\epsilon'\sqrt{K'})\rangle_{\xi_{M,\,M'}}\label{eq:Sbong}
\end{equation}
At this point we can apply the REM--PPP formula for the average \cite{RSBwR,FranchiniSPA2023,Kurkova1}
\begin{equation}
{\textstyle \langle\,\exp\,(\epsilon'\sqrt{K'})\rangle_{\xi_{M,\,M'}}}\overset{d}{=}\exp\left(\lambda K'/2\right)
\end{equation}
and finally find a viable distributional formula for $F$
\begin{equation}
F\left(h_{A}\right)\overset{d}{=}F'+\lambda K'/2
\end{equation}
We have now removed the randomness from our formulas and we can actually
proceed with standard combinatorial methods. For example, the TL can
be deduced as follows: first, notice that $F'$ depends on the rate
parameter $\lambda$ and the two ratios $X/|B|$ and $Y'/|B|$, that
ultimately depend on the magnetization per spin
\begin{equation}
m:=M/|B|.
\end{equation}
In particular, it is shown in \cite{RSBwR,FranchiniSPA2023} that
$\lambda$ depends on the the complexity of $\Phi_{A}\left(\,M,\,M'\right)$.
In this case the complexity is asymptotically the same of $\Omega_{B}\left(\,M,\,M'\right)$
in distribution, so we can study this set directly. As shown in \cite{FranchiniSPA2023},
in the limit of large $N$ is expected that $M'/|B|$ concentrates
on a value that is a function of $m$ only. This, together with the
scaling of $|\,\Omega_{B}\left(\,M\right)|$, allows us to conclude
that the TL of the pressure could be ultimately expressed with a variational
problem in $m$. The exact expressions are given elsewhere \cite{Franchini2024}.
For more informations on the kernel framework see \cite{Franchini2024,RSBwR,FranchiniSPA2023,FranchiniSPA2021,Franchini2017,Franchini2016,Bardella,BFPF}.

\section{Relation with Quantum Field Theory}

We conclude by noticing the similarity with a remarkable exact relation
first introduced by Polyakov, \cite{Polyakov}, between the free energy
of any spin network and the elements of its coupling matrix. The formula
is as follows: consider the uniparted Hamiltonian \cite{Franchini2024}
\begin{equation}
H^{*}\left(\sigma_{A}\right):=\sum_{i\in A}\sum_{j\in A}H_{ij}\sigma_{i}\sigma_{j}
\end{equation}
where the coupling matrix $H$ is assumed arbitrary. We define the
free energy as
\begin{equation}
f^{*}:=\frac{1}{|A|}\log\sum_{\sigma_{A}\in\Omega^{A}}\exp\left(H^{*}\left(\sigma_{A}\right)\right)
\end{equation}
Now, let us assume that $h_{A}$ is some real valued random vector,
distributed according to a multi--variate Gaussian law $\xi$ such
that 
\begin{equation}
\langle\,h_{i}\rangle_{\xi}=0,\ \ \ \langle\,h_{i}^{2}\rangle_{\xi}=\sum_{j\in A}|2H_{ij}|,\ \ \ \langle\,h_{i}h_{j}\rangle_{\xi}=2H_{ij}
\end{equation}
Then, the formula express the free energy in terms of this Gaussian
field:
\begin{equation}
f^{*}=\frac{1}{|A|}\log\,\langle\,\exp\left(F\left(h_{A}\right)\right)\rangle_{\xi}-\frac{1}{|A|}\sum_{i\in A}\left|H_{i\,i}\right|\label{eq:PPK}
\end{equation}
This formula has been first obtained by A. Polyakov (in 1968: Eq.
A.2 of Ref. \cite{Polyakov}), in the framework of Quantum Field Theory
(see the dedicated Chapter 11.1 in the Parisi's book, \cite{Parisi}),
and has been recently rediscovered in the probabilistic context by
Kistler in \cite{Babylon}. The analogy between Eq.s (\ref{eq:frez}),
(\ref{eq:saboom}) and (\ref{eq:PPK}) is evident. See also \cite{Franchini2024,RSBwR,FranchiniSPA2023,FranchiniSPA2021},
for a general theory of uniparted Hamiltonians in the kernel framework.

\section*{Acknowledgments}

This project was partially funded by the European Research Council
(ERC) under European Union\textquoteright s Horizon 2020 research
and innovation programme (Grant agreement No {[}694925{]}). I'm grateful
to Pan Liming (USTC), for important suggestions and for his past efforts
in \cite{BSKold}. I also thank Giorgio Parisi (Accademia dei Lincei),
for bringing to my attention the reference \cite{Polyakov}, and Nicola
Kistler (Goethe University of Frankfurt), for an interesting conversation
about the reference \cite{Babylon}.

\end{document}